\documentclass[twocolumn,aps,prd,superscriptaddress,nofootinbib,preprintnumbers]{revtex4-1}

\usepackage{bm,amsmath,amsfonts,amssymb,graphicx,xspace,placeins,multirow,diagbox}

\usepackage{hyperref}
\usepackage{breakurl}
\usepackage{array}
\usepackage{slashed}
\usepackage{xspace}

\allowdisplaybreaks
\newcommand{\nn}{\nonumber}

\newcommand{\babar}{\mbox{\ensuremath{{\displaystyle B}\!{\scriptstyle A}{\displaystyle B}\!{\scriptstyle AR}}}\xspace}

\def\d{{\rm d}}

\newcommand{\g}{\gamma}

\newcommand{\cbar}{\bar{c}}
\newcommand{\Bbar}{\,\overline{\!B}{}}

\newcommand{\lqcd}{\ensuremath{\Lambda_{\rm QCD}}\xspace}

\newcommand{\ampBb}[2]{\big\langle #1 \big|\, #2\, \big| \Bbar \big\rangle}
\renewcommand{\ampBb}[2]{\langle #1 |\, #2\, | \Bbar \rangle}

\newcommand{\rDs}{r_{D^*}}
\renewcommand{\rDs}{r}

\newcommand\BGLa{\text{BGL$a_1$}\xspace}
\newcommand\BGLb{\text{BGL$b_1$}\xspace}
\newcommand\BGLc{\text{BGL$c_2$}\xspace}

\newcommand\BGLabc[3]{\ensuremath{{\rm BGL}_{{#1}{#2}{#3}}}\xspace}
\renewcommand\BGLa{\BGLabc122}
\renewcommand\BGLb{\BGLabc212}
\renewcommand\BGLc{\BGLabc221}

\renewcommand{\arraystretch}{1.25}
\makeatletter
\g@addto@macro\bfseries{\boldmath}
\makeatother

\usepackage[dvipsnames,table]{xcolor}
\definecolor{red}{rgb}{0.9, 0,0}
\definecolor{blue}{rgb}{0,0.1,0.9}

\begin{document}

\title{$V_{cb}$ Puzzles: The 2018 Edition}
\title{The $V_{cb}$ Files: 2018 Puzzles}
\title{The $V_{cb}$ Files: It is 2018 and we are not alone}
\title{$N = 5$, $6$, $7$, $8$\,: $V_{cb}$? You'll never believe what happens for $N=6$}
\title{$N = 5$, $6$, $7$, $8$\,: 
Nested hypothesis tests and truncation dependence of $|V_{cb}|$}

\author{Florian U.\ Bernlochner}
\affiliation{Karlsruhe Institute of Technology, 76131 Karlsruhe, Germany}

\author{Zoltan Ligeti}
\affiliation{Ernest Orlando Lawrence Berkeley National Laboratory, 
University of California, Berkeley, CA 94720, USA}

\author{Dean J.\ Robinson}
\affiliation{Ernest Orlando Lawrence Berkeley National Laboratory, 
University of California, Berkeley, CA 94720, USA}
\affiliation{Santa Cruz Institute for Particle Physics and
Department of Physics, University of California Santa Cruz,
Santa Cruz, CA 95064, USA}

\begin{abstract}

The determination of $|V_{cb}|$ from exclusive semileptonic $B \to D^*\ell\nu$ decays is sensitive to the choice of form factor parametrization.
Larger $|V_{cb}|$ values are obtained fitting the BGL versus the CLN parametrization to recent Belle measurements. 
For the BGL parametrization, published fits use different numbers of parameters.
We propose a method based on nested hypothesis tests to determine the optimal number of BGL parameters to fit the data, 
and find that six parameters are optimal to fit the Belle tagged and unfolded measurement~\cite{Abdesselam:2017kjf}.
We further explore the differences between fits that use different numbers of parameters.  
The fits which yield $|V_{cb}|$ values in better agreement with determinations from inclusive semileptonic decays,
tend to exhibit tensions with heavy quark symmetry expectations.
These have to be resolved before the determinations of $|V_{cb}|$ from exclusive and inclusive decays can be considered understood.

\end{abstract}

\maketitle

\section{introduction}

In 2017 the Belle Collaboration presented, for the first time, unfolded measurements of the differential decay distributions 
for $\Bbar \to D^* \ell \bar\nu$ decays~\cite{Abdesselam:2017kjf}, and another
measurement appeared more recently~\cite{Abdesselam:2018nnh}.
The unfolded measurement~\cite{Abdesselam:2017kjf} permitted outside groups to perform their own fits to the data, 
using different parametrizations of the $\Bbar \to D^* \ell \bar\nu$ form factors
to extract $|V_{cb}|$.  The choice of form factor parametrizations can have a sizable impact on the extracted value of $|V_{cb}|$.
This is because heavy quark symmetry gives the strongest constraints on the
differential rate at zero recoil (maximal dilepton invariant mass, $q^2$)~\cite{Isgur:1989vq, Isgur:1989ed, Shifman:1987rj, Nussinov:1986hw, Eichten:1989zv, Georgi:1990um, Luke:1990eg, Falk:1990pz}, resulting in both continuum methods and lattice QCD giving the most precise information on the normalization of the rate at zero recoil. 
However, phase space vanishes near maximal $q^2$ as $\sqrt{q^2_{\rm max} - q^2}$, so the measured $q^2$ spectrum has to be fitted over some range to 
extract $|V_{cb}|$. This results in sensitivity to the functional form of the fitted parametrization.

Fitting Belle's unfolded measurement~\cite{Abdesselam:2017kjf} to the BGL parametrization~\cite{Boyd:1995sq, Boyd:1997kz}
yielded higher values of $|V_{cb}|$~\cite{Bigi:2017njr, Grinstein:2017nlq}
than fitting the CLN~\cite{Caprini:1997mu} parametrization
to the same dataset.  (To our knowledge, during 1997--2017 all \babar and Belle measurements of $|V_{cb}|$ from $\Bbar \to D^* \ell\,
\bar\nu$ used the CLN parametrization.)  The BGL results are
in better agreement with $|V_{cb}|$ extracted from inclusive $B\to X_c\ell\bar\nu$ decays~\cite{HFAG}, 
\begin{subequations}\label{fit17}
\begin{align}
|V_{cb}|_{\rm CLN} 		&=  (38.2\pm 1.5)\times 10^{-3}\,,  \quad\,\text{\cite{Abdesselam:2017kjf}}\,, \label{belleVcb}\\
|V_{cb}|_{\BGLabc332} 	& = (41.7^{+2.0}_{-2.1})\times 10^{-3}\,, \qquad \text{\cite{Bigi:2017njr}}\,, \label{gambinoVcb}\\
|V_{cb}|_{\BGLabc222} 	& =  (41.9^{+2.0}_{-1.9})\times 10^{-3}\,, \qquad \text{\cite{Grinstein:2017nlq}}\,. \label{benVcb}
\end{align}
\end{subequations}
Here the \BGLabc{i}{j}{k} notation highlights that these fits have different numbers of parameters (the notation is defined below in Sec.~\ref{sec:FF}), in particular
8 and 6 parameters, respectively. In Ref.~\cite{Abdesselam:2018nnh}, the Belle Collaboration published an ``untagged'' measurement of $\Bbar \to D^* \ell \bar\nu$, 
without fully reconstructing the second $B$ meson in the collision using hadronic decay modes. 
In that analysis, fits to the CLN and a 5-parameter version of the BGL parametrization were performed~\cite{Abdesselam:2018nnh}, and the results are in agreement,
\begin{subequations}\label{fit18}
\begin{align}
|V_{cb}|_{\rm CLN} &= (38.4\pm 0.9)\times 10^{-3}\,, \label{belle18CLN}\\
|V_{cb}|_{\BGLa} &= (38.3\pm 1.0)\times 10^{-3}\,. \label{belle18BGL}
\end{align}
\end{subequations}

The BGL method implements constraints on the shapes of the $B\to D^*$ form factors based on 
analyticity and unitarity~\cite{Bourrely:1980gp, Boyd:1994tt, Boyd:1995cf}.
Three conveniently chosen linear combinations of form factors are expressed in terms of power series in a small conformal parameter, $0 < z \ll 1$.
As indicated in Eqs.~\eqref{fit17} and~\eqref{fit18}, there are varying choices for the total number of coefficients, $N$, in the three power series, 
ranging from $N=5$~\cite{Abdesselam:2018nnh}, to $N=6$~\cite{Grinstein:2017nlq, Bernlochner:2017xyx}, and $N=8$~\cite{Bigi:2017njr, Bigi:2017jbd, Jaiswal:2017rve}.
The CLN~\cite{Caprini:1997mu} prescription uses similar analyticity and unitarity constraints on the $B \to D$ form factor,
heavy quark effective theory (HQET)~\cite{Georgi:1990um, Eichten:1989zv} relations between the $B\to D$ and $B\to D^*$ form factors, 
and QCD sum rule calculations~\cite{Neubert:1992wq, Neubert:1992pn, Ligeti:1993hw} of the order $\lqcd/m_{c,b}$ subleading Isgur-Wise functions~\cite{Luke:1990eg, Falk:1990pz}. 
It has 4 fit parameters. (This version of the CLN parametrization, as used to extract $|V_{cb}|$, is not self consistent at $\mathcal{O}(\lqcd/m_{c,b})$~\cite{Bernlochner:2017jka}.)

The relation between the above fits is nontrivial, and has not been studied systematically.  The goal of this paper is to explore their differences,
and to devise a quantitative method to identify the optimal number of parameters in the BGL framework. Using a prescription based on a nested hypothesis test, we find that at least $6$ parameters are required to describe the data from Ref.~\cite{Abdesselam:2017kjf}.
The $N=5$ and 6 fits we study in detail, yield $|V_{cb}|$ values in better agreement with determinations from inclusive semileptonic decays, 
but exhibit tensions with expectations from heavy quark symmetry.

\section{Formalism and notations}
\label{sec:FF}

The vector and axial-vector $\Bbar \to D^*$ form factors are defined as
\begin{align}\label{formfactors}
\ampBb{D^*}{\cbar \g^\mu b} & = i\sqrt{m_B m_{D^*}}\, h_V\, 
  \varepsilon^{\mu\nu\alpha\beta}\,
  \epsilon^*_{\nu}v'_\alpha v_\beta \,, \nn\\*
\ampBb{D^*}{\cbar \g^\mu \g^5 b} & = \sqrt{m_B m_{D^*}}\, 
  \big[h_{A_1} (w+1)\epsilon^{*\mu} \\*
  & \quad - h_{A_2}(\epsilon^* \cdot v)v^\mu 
  - h_{A_3}(\epsilon^* \cdot v)v'^\mu \big] , \nn
\end{align}
where $v$ ($v'$) is the four-velocity of the $B$ ($D^{*}$). The
form factors $h_{V,A_{1,2,3}}$ depend on $w = v\cdot v' =
(m_B^2+m_{D^{*}}^2-q^2) / (2m_B m_{D^{*}})$.  In the heavy quark limit, $h_{A_1}
= h_{A_3} = h_V =  \xi$ and $h_{A_2} = 0$, where $\xi$ is the Isgur-Wise
function~\cite{Isgur:1989vq, Isgur:1989ed}.  Each of these form factors can be
expanded in powers of $\lqcd / m_{c,b}$ and $\alpha_s$.  

In the massless lepton limit (i.e., $\ell = e$ or $\mu$), the differential $B\to D^*\ell\bar\nu$ rate is given by
\begin{align}\label{dGdw}
\frac{\d \Gamma}{\d w} & = \frac{G_F^2|V_{cb}|^2\, \eta_{\rm ew}^2\, m_B^5}{48 \pi^3}\,
  \sqrt{w^2-1}\, (w + 1)^2\, \rDs^3 (1- \rDs)^2 \nn \\*
  & \qquad \times \bigg[1 + \frac{4w}{w+1}
  \frac{1- 2 w\rDs + \rDs^2}{(1 - \rDs)^2} \bigg] [\mathcal{F}(w)]^2\,,
\end{align}
where $\rDs = m_{D^*}/m_B$, and ${\cal F}(w)$ can be written in terms of $h_{A_1}(w)$ and the two form factor ratios (see, e.g., Ref.~\cite{Manohar:2000dt})
\begin{equation}\label{eqn:R1R2Def}
  R_1(w) = \frac{h_V}{h_{A_1}}\,, \qquad 
  R_2(w) = \frac{h_{A_3} + \rDs\, h_{A_2}}{h_{A_1}}\,.
\end{equation}
All measurable information is then contained in the three functions ${\cal F}(w)$ and $R_{1,2}(w)$.
Throughout this paper,  ${\cal F}(1) = 0.906$~\cite{Bailey:2014tva} and $\eta_{\rm ew} = 1.0066$~\cite{Sirlin:1981ie}
are used to convert fit results for $|V_{cb}|\, {\cal F}(1)\, \eta_{\rm ew}$ to
values of $|V_{cb}|$.	
In the heavy quark limit $R_{1,2}(w) = 1 + {\cal O}(\lqcd/m_{c,b},\, \alpha_s)$ and ${\cal F}(w) = \xi(w)$. 
Thus, $R_{1,2}(w)-1$ parametrize deviations from the heavy quark limit.

The BGL framework is defined by expanding three form factors $g$, $f$, and ${\cal F}_1$, which are linear combinations of those defined in Eq.~(\ref{formfactors}), in power series of the form $1/[ P_i(z)\phi_i(z) ] \times \sum a_n^i z^n$, where $i = g$, $f$, ${\cal F}_1$ (see, e.g., Ref.~\cite{Boyd:1997kz}, and note that ${\cal F}_1 \neq {\cal F}$). 
Here $z=z(w)$ is a conformal parameter that maps the physical region $1 < w < 1.5$ onto $0<z<0.056$, 
and $P_i(z)$ and $\phi_i(z)$ are known functions~\cite{Grinstein:2017nlq}. 
There are two notations in the literature for the coefficients of these power series, which map onto each other via
\begin{equation}
\label{eq:bglff}
\big\{ a_n,\, b_n,\, c_n \big\} \ \text{\cite{Grinstein:2017nlq}}\ 
  \longleftrightarrow \ \big\{ a_n^g,\, a_n^f,\,
  a_n^{{\mathcal F_1}} \big\} \ \text{\cite{Bigi:2017njr}} \,.
\end{equation}
In the remainder of this paper we adopt the former notation, so that $a_n$, $b_n$ and $c_n$ are the coefficients of $g$, $f$, and ${\cal F}_1$, respectively.
(The convention 
for the sign of $g$, and thus the $a_n$, in Ref.~\cite{Grinstein:2017nlq} is opposite to that used in Refs.~\cite{Bigi:2017njr, Jaiswal:2017rve}.)
Note that $c_0$ is fixed by $b_0$~\cite{Boyd:1997kz, Grinstein:2017nlq}, and the fits are performed for the rescaled parameters
\begin{equation}\label{tildeparam}
\big\{ \tilde{a}_n, \tilde{b}_n, \tilde{c}_n\big\} 
  = \eta_{\rm ew}\, |V_{cb}|\, \big\{ a_n, b_n, c_n\big\}\,,
\end{equation}
and $|V_{cb}|$ is determined by $|\tilde b_0|$.

To study and distinguish expansions truncated at different orders in $z$, we
denote by \BGLabc{n_a}{n_b}{n_c} a BGL fit with the parameters,
\begin{equation}\label{BGLabcdef}
\{ a_{0,\ldots,\, n_a-1},\, b_{0,\ldots,\, n_b-1},\, c_{1,\ldots,\, n_c}\}\,.
\end{equation} 
The total number of fit parameters is $N = n_a + n_b + n_c$.  The BGL parametrization used in Refs.~\cite{Grinstein:2017nlq, Bernlochner:2017xyx}, is \BGLabc222, while that used in Refs.~\cite{Bigi:2017njr, Jaiswal:2017rve} is \BGLabc332.

\begin{table*}[tb]
\renewcommand{\arraystretch}{1.25}
\newcolumntype{C}{ >{\centering\arraybackslash $} m{1.5cm} <{$}}
\begin{tabular}{c|CCC|CCC|CCC} 
\hline\hline
\diagbox{~$n_c\!\!$}{$\!\!n_a$~} & 1 & 2 & 3 & 1 & 2 & 3 & 1 & 2 & 3 \\
\hline
\multirow{2}{*}{1}      &  33.2 & 31.6	& \cellcolor{blue!10}31.2	&  33.0	& \cellcolor{blue!10}29.1 	
			&  \cellcolor{green!10}28.9 		&  \cellcolor{blue!10}30.4 	& \cellcolor{green!10}29.1 	& \cellcolor{orange!10}28.9 \\
			&  38.6 \pm 1.0 & 38.6 \pm 1.0 	& \cellcolor{blue!10}38.6 \pm 1.0 	&  39.0 \pm 1.5 
			& \cellcolor{blue!10}40.7 \pm 1.6 	& \cellcolor{green!10}40.7 \pm 1.6 		&  \cellcolor{blue!10}40.7 \pm 1.7 	& \cellcolor{green!10}40.6 \pm 1.8 	& \cellcolor{orange!10}40.6 \pm 1.8
\\ &&&&&&\\[-8pt]
\multirow{2}{*}{2} 	&  32.9 & \cellcolor{blue!10}31.3	& \cellcolor{green!10}31.1	&  \cellcolor{blue!10}32.7	& \cellcolor{green!10}\bf{27.7} 	
			& \cellcolor{orange!10}27.7	&  \cellcolor{green!10}29.2	& \cellcolor{orange!10}27.7 	& \cellcolor{red!10}27.7\\ 
			& 38.8 \pm 1.1 	& \cellcolor{blue!10}38.7 \pm 1.1 	&  \cellcolor{green!10}38.8 \pm 1.0 & \cellcolor{blue!10}39.5 \pm 1.7 	&\cellcolor{green!10} \bm{41.7 \pm 1.8} 	
			&  \cellcolor{orange!10}41.6 \pm 1.8 	& \cellcolor{green!10}41.8 \pm 2.0 	& \cellcolor{orange!10}41.8 \pm 2.0 & \cellcolor{red!10}41.7 \pm 2.0
\\ &&&&&&\\[-8pt]	
\multirow{2}{*}{3} 	&  \cellcolor{blue!10}31.7 	& \cellcolor{green!10}31.3	& \cellcolor{orange!10}31.0 &  \cellcolor{green!10}29.1 		
			& \cellcolor{orange!10}27.7  &  \cellcolor{red!10}27.6 		&  \cellcolor{orange!10}29.2 	& \cellcolor{red!10}27.6 	& 23.2\\
			& \cellcolor{blue!10}39.0 \pm 1.1 	& \cellcolor{green!10}38.6 \pm 1.2 	& \cellcolor{orange!10}38.6 \pm 1.1  
			&  \cellcolor{green!10}41.9 \pm 2.0 & \cellcolor{orange!10}41.8 \pm 2.0 		& \cellcolor{red!10}41.7 \pm 2.0 	&  \cellcolor{orange!10}41.8 \pm 2.0 	& \cellcolor{red!10}41.7 \pm 1.9 	& 41.4 \pm 2.0
\\ \hline	
& \multicolumn{3}{c|}{ $n_b = 1$ } & \multicolumn{3}{c|}{ $n_b = 2$ } & \multicolumn{3}{c}{ $n_b = 3$ } \\
\hline\hline
\end{tabular}
\caption{The $\chi^2$ (upper entry) and $|V_{cb}| \times 10^3$ (lower entry) values for the \BGLabc{n_a}{n_b}{n_c} fits used for the nested hypothesis test.  The number of free parameters in a given
fit is $N = n_a + n_b + n_c$ and the bold entry is the
selected \BGLabc222 hypothesis $\{a_0,\, a_1,\, b_0,\, b_1,\, c_1,\, c_2\}$. Cells corresponding to $N =
5$, $6$, $7$, $8$ are highlighted blue, green, orange, and red, respectively. }
\label{tab:chi2Vcb}
\end{table*}

\section{Nested Hypothesis Tests: fixing the optimal number of coefficients}
\label{sec:NHT}

Our aim is to construct a prescription to determine the optimal number of parameters to fit a given data set. 
This can be achieved by use of a nested hypothesis test: a test of an $N$-parameter fit hypothesis versus a fit using one additional parameter (the alternative hypothesis).

Such a hypothesis test requires an appropriate statistical measure or test statistic. A suitable choice is the difference in $\chi^2$,
\begin{align}
 \Delta \chi^2 = \chi^2_{N} - \chi^2_{N+1} \,.
\end{align} 
The fit with one additional parameter --- the $(N+1)$-parameter fit --- has one fewer degree of freedom (number of bins minus the number of parameters). 
In the large number of degrees of freedom limit, $\Delta \chi^2$ is distributed as a $\chi^2$ with a single degree of freedom~\cite{wilks1938}.
One may reject or accept the alternative hypothesis by choosing a decision boundary.  
If, for instance, we choose $\Delta \chi^2 = 1$ as the decision boundary, we would reject the $(N+1)$-parameter hypothesis in favor 
of the $N$-parameter fit 68\% of the time, if the $N$ parameter hypothesis is true.

We seek a prescription to incrementally apply this nested hypothesis test, starting from a suitably small initial number of parameters
(to avoid possible overfitting), until we reach the simplest (smallest $N$) fit containing the initial parameters, that is preferred over all
hypotheses that nest it or are nested by it. For a set of BGL fits, we thus propose the following prescription starting from a suitable low-$N$ fit \BGLabc{n_a}{n_b}{n_c}:
\begin{itemize}\itemsep 0pt
\item[(i)] Carry out fits with one parameter added (a ``descendant'' fit) or, when permitted, removed (a ``parent'' fit);
i.e., for \BGLabc{(n_a\pm 1)}{n_b}{n_c}, \BGLabc{n_a}{(n_b\pm 1)}{n_c}, \BGLabc{n_a}{n_b}{(n_c\pm 1)}.
\item[(ii)] For each descendant (parent) hypothesis, accept it over \BGLabc{n_a}{n_b}{n_c} if $\Delta \chi^2$ is above (below) the decision boundary value.
\item[(iii)] Repeat (i) and (ii) recursively, until a ``stationary'' fit is reached, that is preferred over its parents and descendants.
\item[(iv)] If there are multiple stationary fits, choose the one with the smallest $N$, then the smallest $\chi^2$.
\end{itemize}
The optimal truncation order obtained this way depends on the precision of the available experimental data. Our prescription attempts to minimize the residual model dependence (caused by this truncation) with respect to the experimental uncertainty. 

Table~\ref{tab:chi2Vcb} shows the fitted $\chi^2$ values for the set of 27 different \BGLabc{n_a}{n_b}{n_c} fits with $n_i = 1,2,3$. 
A suitable choice for a starting fit is \BGLabc111 or one of the three possible fits with $N = 4$.
Using the decision boundary of $\Delta \chi^2 > 1$, one then obtains a single stationary solution, \BGLabc222 shown in bold. For example, one path to \BGLabc222 is $111 \to 211 \to 221 \to 222$, while another is $121 \to 131 \to 231 \to 232 \to 222$. 

Also shown in Table~\ref{tab:chi2Vcb} are the $|V_{cb}|$ values for all 27 fits. These results are consistent with the statement made in Ref.~\cite{Bigi:2017njr} that the extracted values of $|V_{cb}|$ remain stable when one adds more fit parameters to the \BGLabc332 fit. This stability can be seen directly by comparing the preferred \BGLabc222 fit with its descendants.  One may notice that the $\chi^2$ of the \BGLabc333 fit is  substantially smaller than those of its parents.  However, our procedure starting from the $N=3$ or 4 fits always terminates before reaching so many parameters.  Plotting the fitted \BGLabc333 distributions, one sees that its small $\chi^2$ is due to fitting fluctuations in the data, and should be seen as an overfit.

The unitarity constraints, $\sum_{n=0}^\infty |a_n|^2 \leq 1$ and 
$\sum_{n=0}^\infty \big( |b_n|^2 + |c_n|^2 \big) \leq 1$, can be imposed on the fits.  The stationary fit in our approach, \BGLabc222, is far from saturating these bounds~\cite{Grinstein:2017nlq}.  
While the form factors must obey the unitarity constraints, statistical fluctuations in their binned measurements may cause the central values to appear to violate unitarity\footnote{We thank Paolo Gambino for raising this question.} (at a modest confidence level).  This can occur because such fits may yield large coefficients for higher order terms to accommodate ``wiggles'' in the data.  In this paper we do not impose unitarity as a constraint; fits whose central values violate unitarity (at a modest confidence level) may suggest an overfit.
This is the case for the \BGLabc333 fit, providing another reason to limit the number of fit coefficients, as proposed in our method.

\section{Comparing $N=5$ fits with \BGLabc222}
\label{sec:5fits}

To explore the differences between the various $5$-parameter fits and the \BGLabc222 fit, 
we perform such fits to Belle's unfolded data~\cite{Abdesselam:2017kjf}.  
(The untagged Belle measurement~\cite{Abdesselam:2018nnh} is not unfolded, and cannot be analyzed at this point outside the Belle framework.
With limited statistics, the differences between the fits we perform on the unfolded data contain fluctuations, which are different from those of the folded measurement.)
There are six possible fits with 5 parameters, as shown in Table~\ref{tab:chi2Vcb}.  
Here we focus on comparing \BGLa, \BGLb, \BGLc, which respectively set $a_1$, $b_1$, or $c_2$ to zero. 
(We do not study further the \BGLabc311, \BGLabc131, and \BGLabc113 fits, as each removes two and adds one parameter to the \BGLabc222~fit.)

The results of the \BGLabc222 fit and the three 5-parameter fits for the physical observables 
$|V_{cb}|$, $R_{1,2}(1)$, and $R_{1,2}'(1)$ are shown in Table~\ref{tab:compare}.  
(Our \BGLabc222 fit results vary slightly from those in Ref.~\cite{Bernlochner:2017xyx}, due to using $m_B = 5.280$\,GeV versus $5.279$\,GeV.)
The best fit parameters [rescaled as in Eq.~(\ref{tildeparam})] and correlations for these four fits are shown in Table~\ref{tab:BGLtag}.

\begin{table}[b]\tabcolsep 3pt
\resizebox{\linewidth}{!}{
\begin{tabular}{c|cccc}
\hline\hline
&  \BGLabc222 & \BGLa & \BGLb & \BGLc \\
\hline  
$\chi^2 / \!$ ndf	&  27.7/34  &  32.7/35  &  31.3/35  &  29.1/35  \\
$|V_{cb}| \!\times\! 10^3$  &  $41.7 \pm 1.8$  &  $39.5 \pm 1.7$
  &  $38.7 \pm 1.1$  &  $40.7 \pm 1.6$  \\
\hline
$R_1(1)$	&  $0.45 \pm 0.31$ & $1.30 \pm 0.09$ & $0.86 \pm 0.37$ & $0.48 \pm 0.34$  \\
$R_1'(1)$	&  $4.23 \pm 1.28$ & $0.26 \pm 0.27$ & $2.34 \pm 1.60$ & $4.02 \pm 1.44$  \\
$R_2(1)$	&  $1.00 \pm 0.19$ & $1.03 \pm 0.20$ & $1.05 \pm 0.20$ & $0.82 \pm 0.10$  \\
$R_2'(1)$	&  $-0.53 \pm 0.43$ & $-0.29 \pm 0.51$ & $-0.25 \pm 0.52$ & $-0.02 \pm 0.05$  \\
\hline\hline
\end{tabular}}
\caption{Summary of the \BGLabc222, \BGLa, \BGLb, and \BGLc fits to the
tagged and unfolded Belle measurement~\cite{Abdesselam:2017kjf}.}
\label{tab:compare}
\end{table}

\begin{table*}[t!]
\begin{tabular}{ccr|rrrrrr} 
\hline \hline
\multirow{8}{*}{\BGLabc222} & \multirow{2}{*}{Param}  &  \multirow{2}{*}{Value $\times\ 10^2$~~}  
  &  \multicolumn{6}{c}{Correlation} \\
& &  &  $\tilde a_0$  &  $\tilde a_1$   &  $\tilde b_0$	& $\tilde b_1$ & $\tilde c_1$ & $\tilde c_2$ \\ \cline{2-9}
&$\tilde a_0$  &  $0.0379 \pm 0.0249$  &  1.000  &  $-$0.952  &  $-$0.249  &  0.417  &  0.137  &  $-$0.054\\
&$\tilde a_1$  &  $2.6954 \pm 0.9320$  &  &  1.000  &  0.383  &  $-$0.543  &  $-$0.268  &  0.165\\
&$\tilde b_0$  &  $0.0550 \pm 0.0023$  &  &  &  1.000  &  $-$0.793  &  $-$0.648  &  0.461\\
&$\tilde b_1$  &  $-0.2040 \pm 0.1064$ &  &  &  &  1.000  &  0.542  &  $-$0.333\\
&$\tilde c_1$  &  $-0.0433 \pm 0.0264$ &  &  &  &  &  1.000  &  $-$0.953\\
&$\tilde c_2$  &  $0.5350 \pm 0.4606$  &  &  &  &  &  &  1.000\\
\hline\hline
\multirow{7}{*}{\BGLabc122} & \multirow{2}{*}{Param}  &  \multirow{2}{*}{Value $\times\ 10^2$~~}  
  &  \multicolumn{6}{c}{Correlation} \\
& &  &  $\tilde a_0$  &  $\tilde b_0$	& $\tilde b_1$ & $\tilde c_1$ & $\tilde c_2$  &\\ \cline{2-9}
&$\tilde a_0$  &  $0.1066 \pm 0.0070$	&  1.000  &  0.271  &  $-$0.163  &  $-$0.316  &  0.297 & \\
& $\tilde b_0$  &  $0.0521 \pm 0.0022$	&  &  1.000  &  $-$0.767  &  $-$0.612  &  0.432 &\\
& $\tilde b_1$  &  $-0.0446 \pm 0.0839$	&  &  &  1.000  &  0.489  &  $-$0.287 &\\
& $\tilde c_1$  &  $-0.0193 \pm 0.0252$	&  &  &  &  1.000  &  $-$0.956 &\\
& $\tilde c_2$  &  $0.2654 \pm 0.4492$	&  &  &  &  &  1.000 & \\
\hline\hline
\multirow{7}{*}{\BGLabc212} & \multirow{2}{*}{Param}  &  \multirow{2}{*}{Value $\times\ 10^2$~~}  
  &  \multicolumn{6}{c}{Correlation} \\
& &  &  $\tilde a_0$  &  $\tilde a_1$   &  $\tilde b_0$ & $\tilde c_1$ & $\tilde c_2$ & \\ \cline{2-9}
& $\tilde a_0$  &  $0.0672 \pm 0.0288$  &  1.000  &  $-$0.972  &  0.128  &  $-$0.061  &   0.053 & \\
& $\tilde a_1$  &  $1.4254 \pm 1.0155$  &  &  1.000  &  $-$0.074  &  $-$0.005  &   0.010 & \\
& $\tilde b_0$  &  $0.0511 \pm 0.0014$  &  &  &  1.000  &  $-$0.420  &  0.342 & \\
& $\tilde c_1$  &  $-0.0140 \pm 0.0223$ &  &  &  &  1.000  &  $-$0.976 & \\
& $\tilde c_2$  &  $0.2187 \pm 0.4367$  &  &  &  &  &  1.000 & \\
\hline\hline
\multirow{7}{*}{\BGLabc221} &\multirow{2}{*}{Param}  &  \multirow{2}{*}{Value $\times\ 10^2$~~}  
  &  \multicolumn{6}{c}{Correlation} \\
& &  &  $\tilde a_0$  &  $\tilde a_1$	& $\tilde b_1$ & $\tilde b_1$ & $\tilde c_1$ & \\ \cline {2-9}
& $\tilde a_0$  &  $0.0399 \pm 0.0270$  &    1.000  &  $-$0.965  &  $-$0.294  &  0.472  &  0.330 \\
& $\tilde a_1$  &  $ 2.5020 \pm 0.9984$  &  &  1.000  &  0.380  &  $-$0.555  &  $-$0.408\\
& $\tilde b_0$  &  $0.0537 \pm 0.0021$  &  &  &  1.000  &  $-$0.774  &  $-$0.787\\
& $\tilde b_1$  &  $-0.1618 \pm 0.1020$  &  &  &  &  1.000  &  0.799\\
& $\tilde c_1$  &  $-0.0141 \pm 0.0082$  &  &  &  &  &  1.000\\
\hline\hline
\end{tabular}
\caption{Fit coefficients and correlation matrices for the 6-parameter \BGLabc222 fit and three 5-parameter BGL fits to the
tagged and unfolded Belle measurement~\cite{Abdesselam:2017kjf}.}
\label{tab:BGLtag}
\end{table*}

The results for the \BGLabc222 fit in Table~\ref{tab:BGLtag} suggest that, if one wants to
reduce the number of fit parameters from 6 to 5, 
the \BGLabc122 fit might be the least optimal choice, 
as the significance of a nonzero value for $|a_1|$ is greater than for $|b_1|$, 
which is turn greater than for $|c_2|$.
This is in line with the observation that, compared to the \BGLabc222 fit, the value of $\chi^2$ increases the most
for \BGLa, followed by \BGLb, and then \BGLc. 
This suggests that among the 5-parameter fits setting $c_2 = 0$ (the \BGLabc221 fit) may instead be the preferred option,
though inferior, according to our method, to the \BGLabc222 fit for the Belle tagged and unfolded dataset~\cite{Abdesselam:2017kjf}.

\begin{figure*}[t]
  \includegraphics[width=0.45\textwidth]{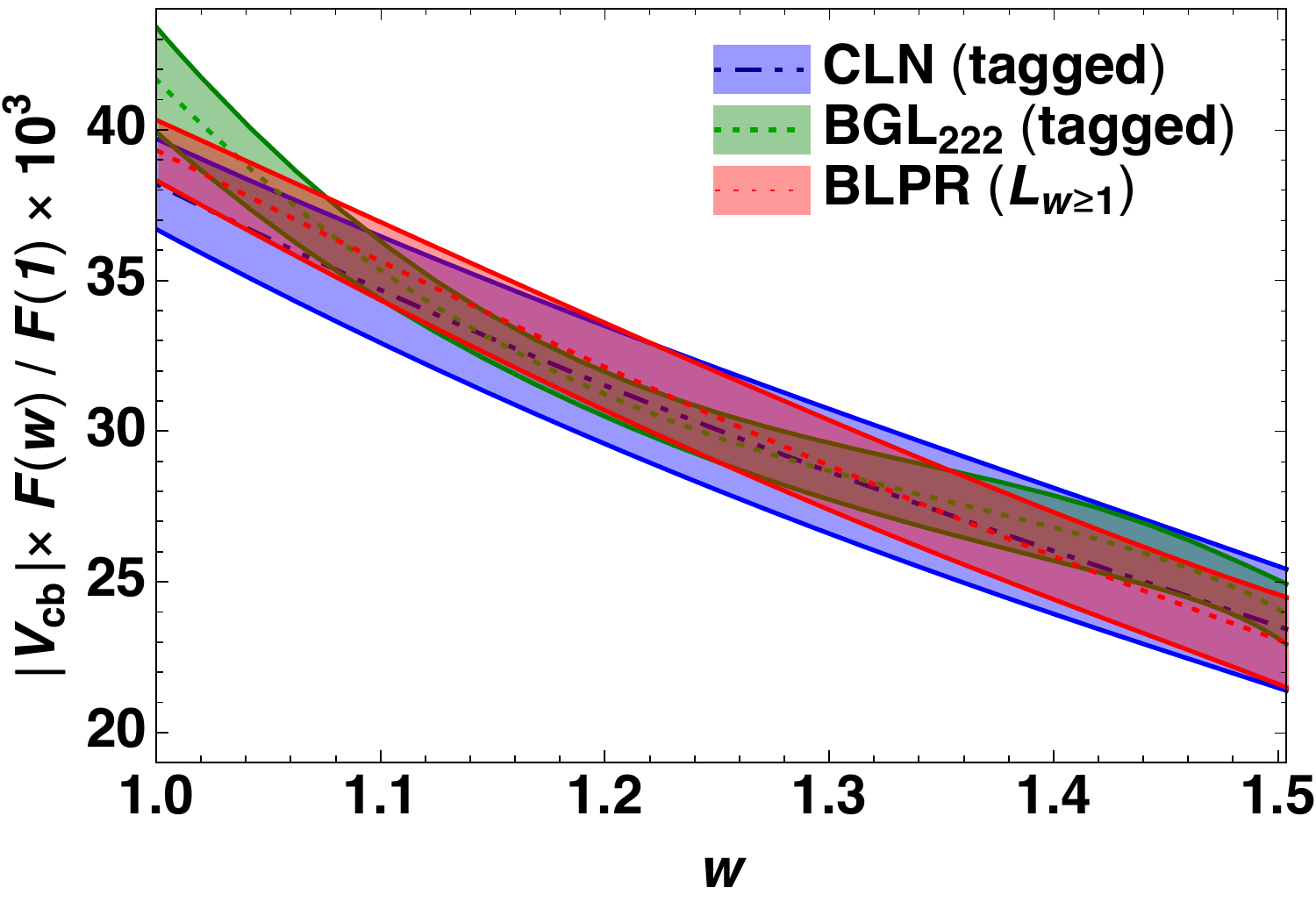} \hfil
  \includegraphics[width=0.45\textwidth]{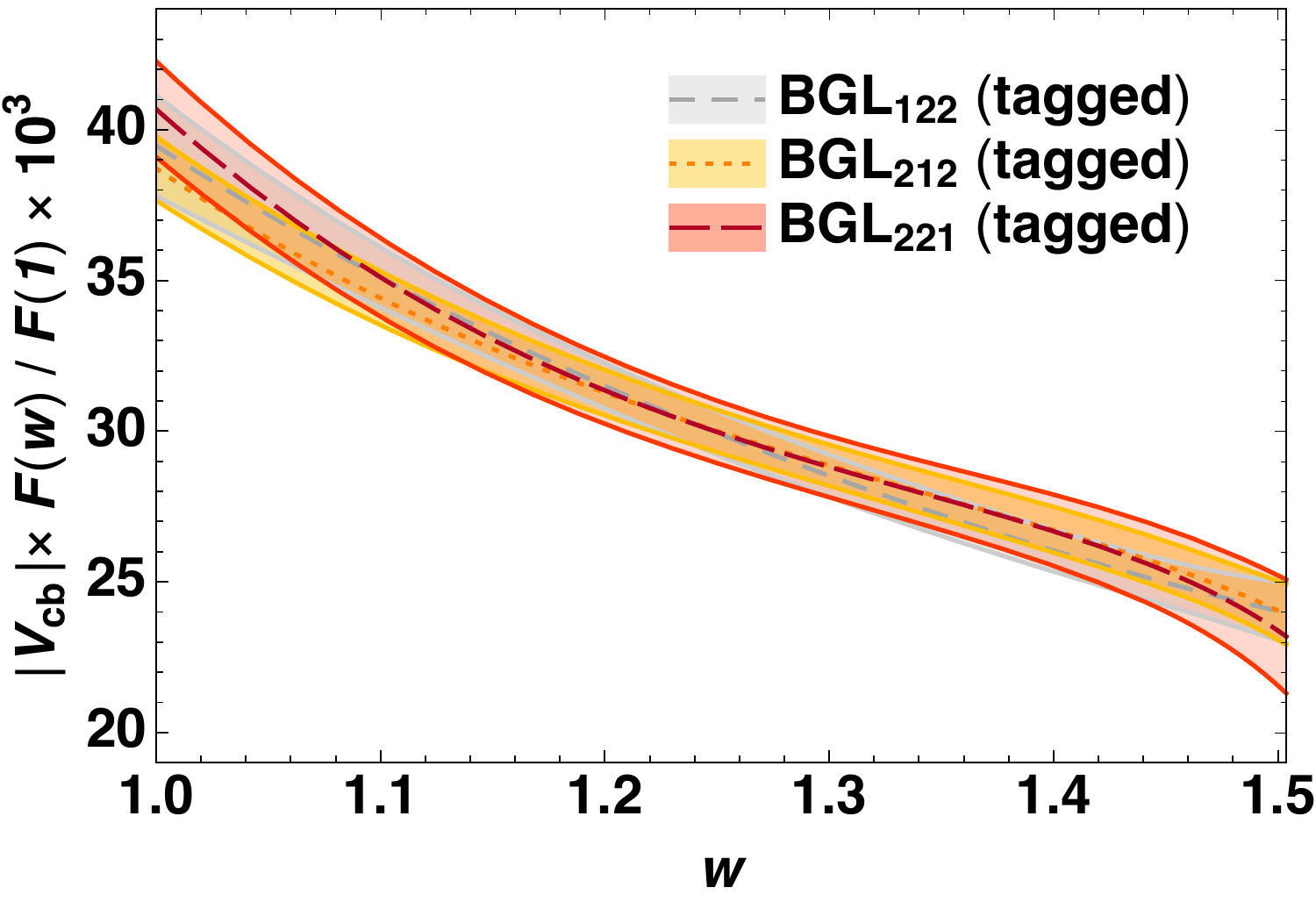} \\
  \includegraphics[width=0.45\textwidth]{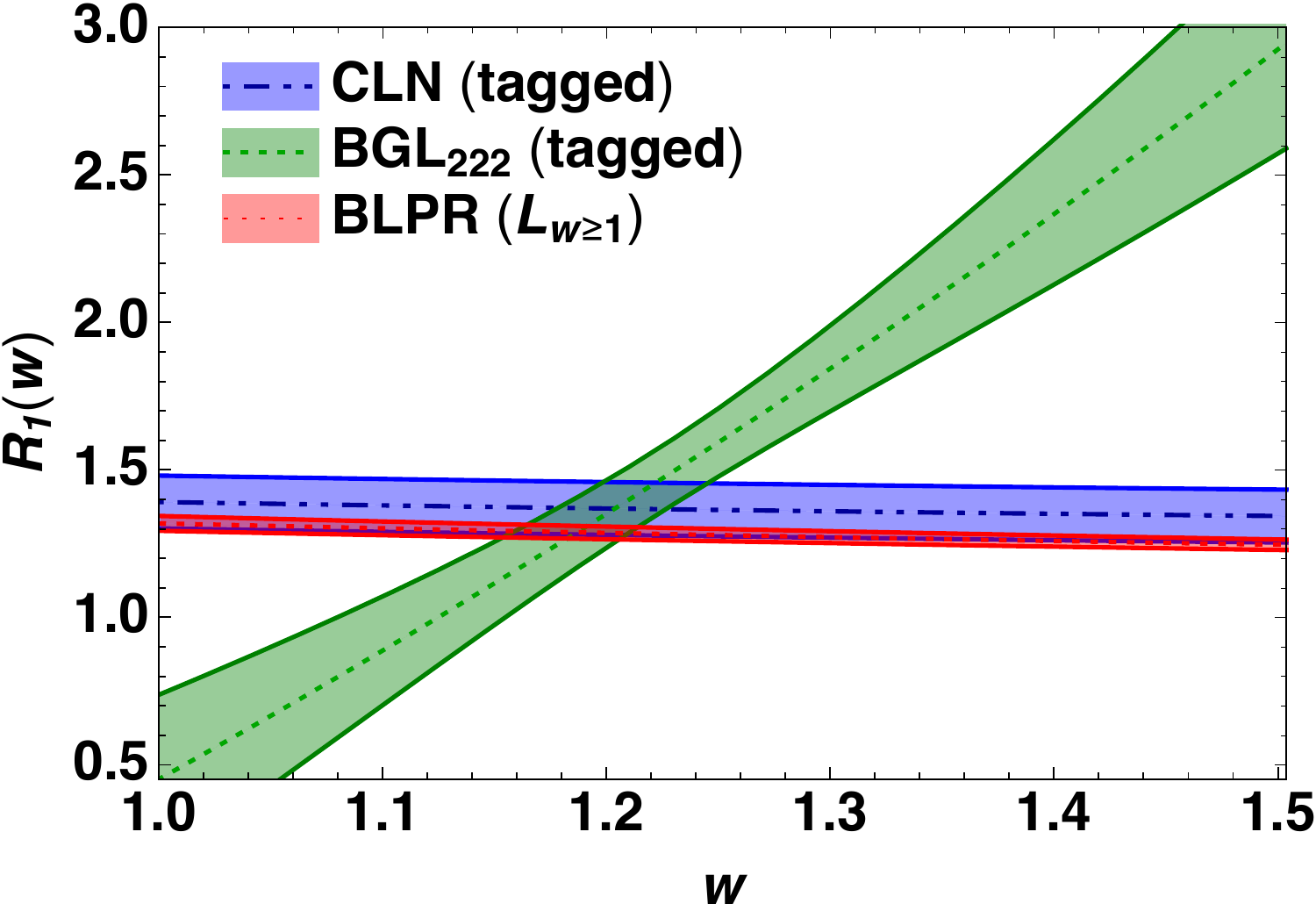} \hfil
  \includegraphics[width=0.45\textwidth]{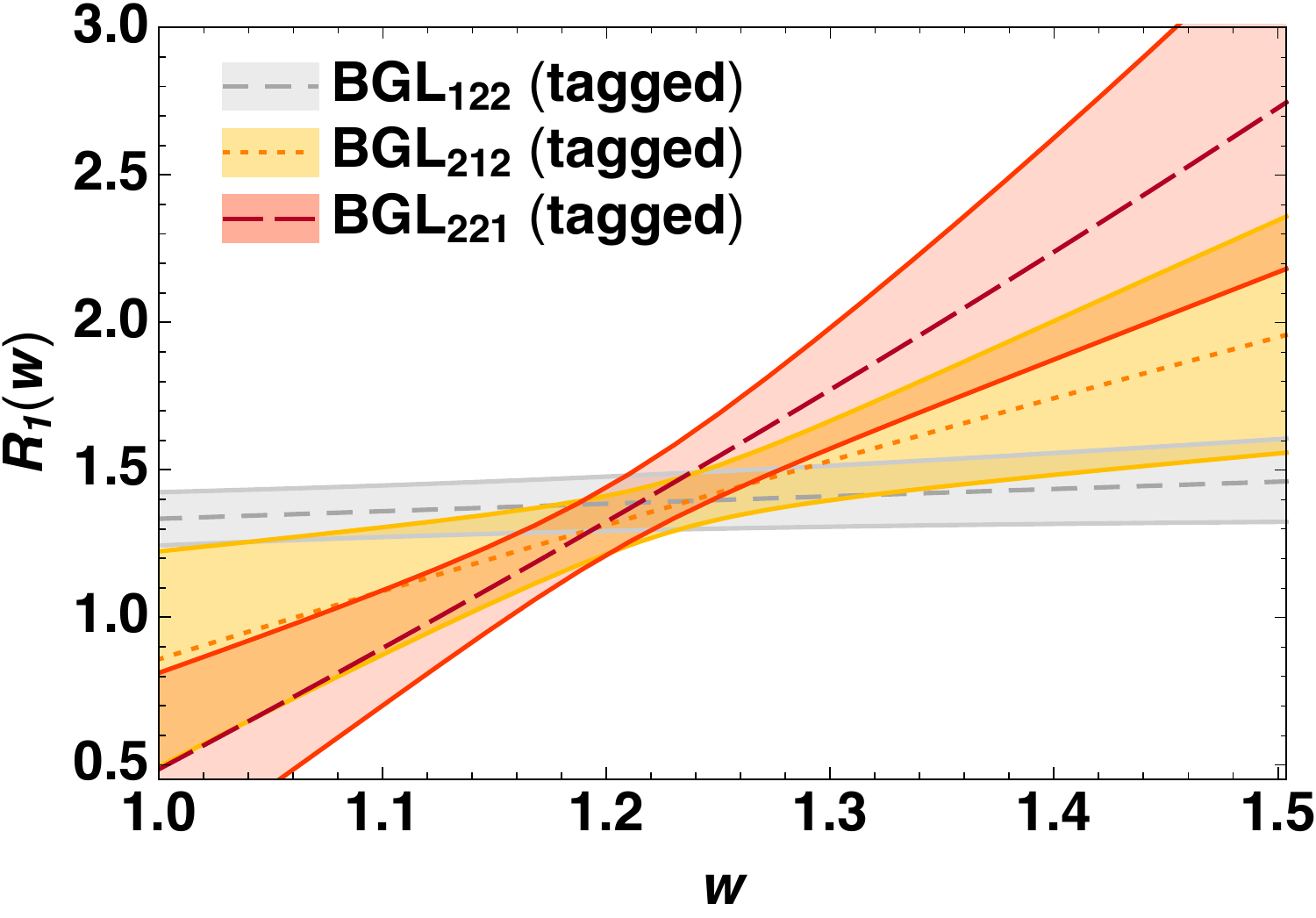} \\
  \includegraphics[width=0.45\textwidth]{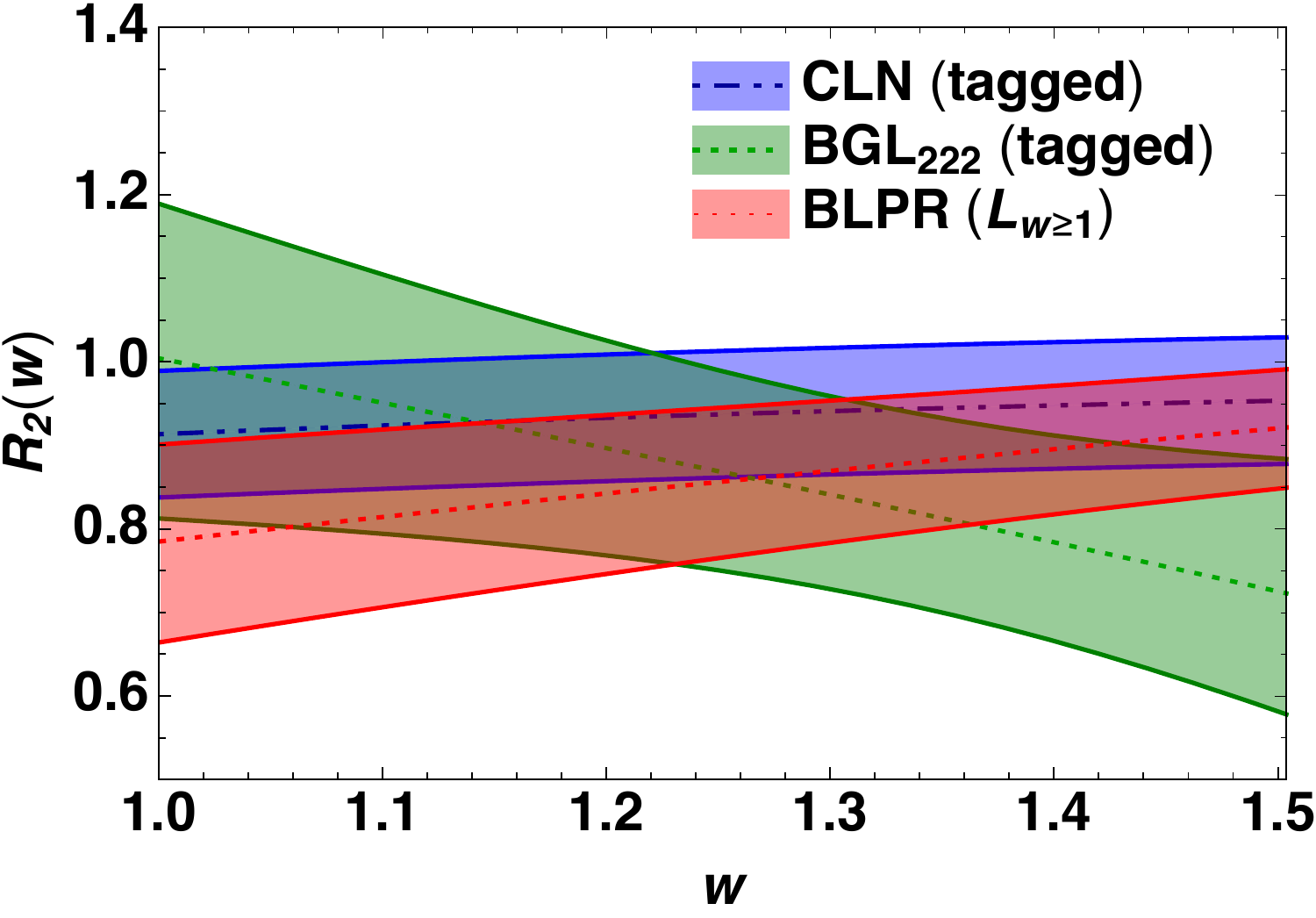} \hfil
  \includegraphics[width=0.45\textwidth]{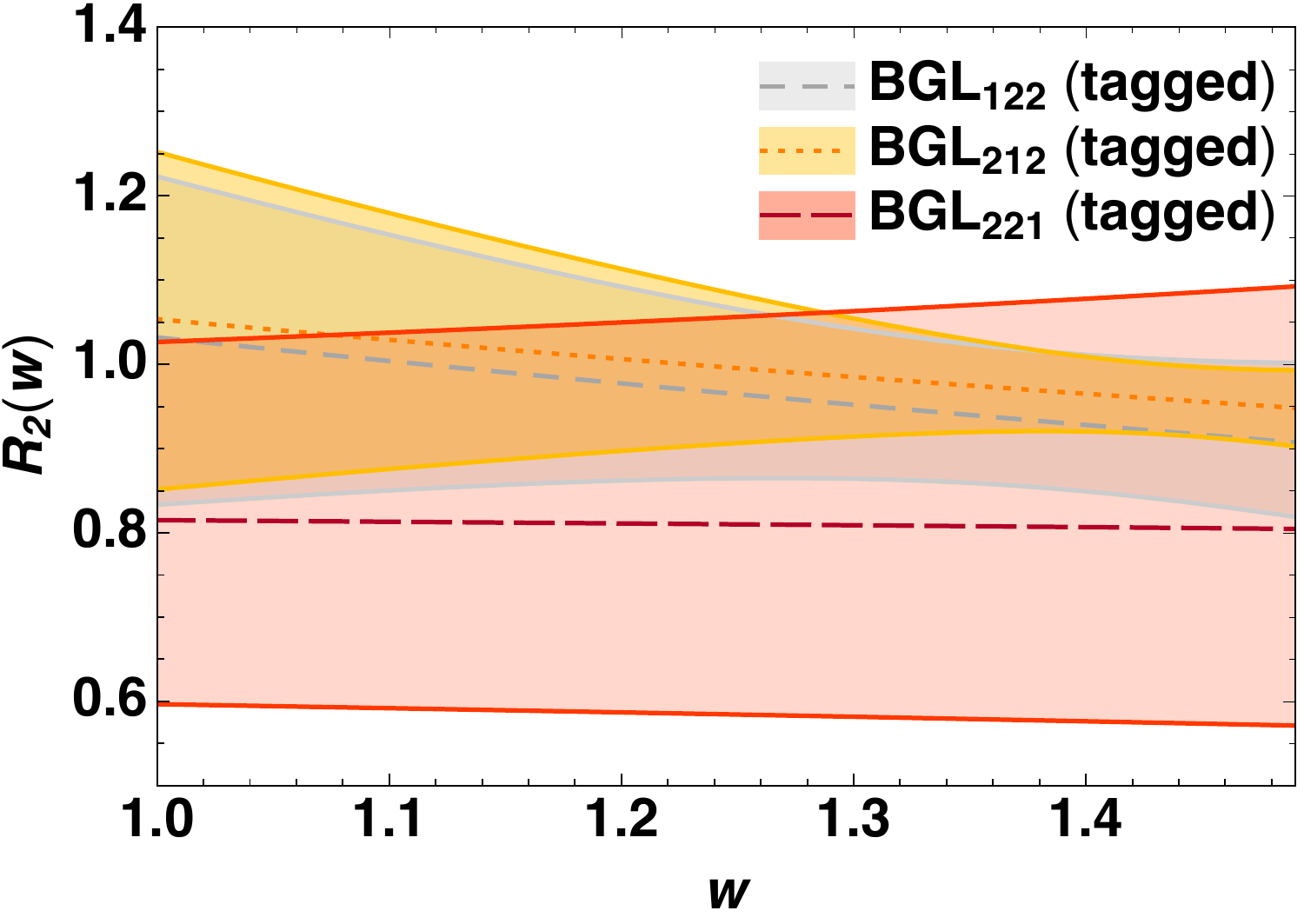}
\caption{The form factor ${\cal F}(w)$ (top), $R_1(w)$ (middle) and $R_2(w)$ (bottom) for the six fits described in
the text.}
\label{fig:FR12}
\end{figure*}

The top row in Fig.~\ref{fig:FR12} shows ${\cal F}(w)$ normalized to the lattice QCD value of
${\cal F}(1)$, as $|V_{cb}|\, {\cal F}(w) / {\cal F}(1)$ for six fits.
The left-side plots show three previously published fits: the \BGLabc222 and CLN fit results, based on the 2017 Belle tagged measurement,  
and the `BLPR' result of Ref.~\cite{Bernlochner:2017jka}, 
which performed an HQET-based fit to both $B\to D^* l\bar\nu$ and $B \to D l\bar\nu$ data to determine the subleading ${\cal O}(\lqcd/m_{c,b})$ Isgur-Wise functions, using also lattice QCD information.  
The right-side plots in Fig.~\ref{fig:FR12} show the \BGLa,
\BGLb, and \BGLc fits, based on the 2017 Belle tagged
measurement~\cite{Abdesselam:2017kjf}.  The shaded bands indicate the
uncertainties.  The \BGLabc222 and \BGLc fits have the largest differential rates near
zero recoil ($w=1$), corresponding to the largest extracted values of
$|V_{cb}|$.

The value of $|V_{cb}|$ extracted from the \BGLa fit to the 2017 Belle unfolded
measurement~\cite{Abdesselam:2017kjf} is more than $1\sigma$ smaller than in the 6-parameter \BGLabc222 fit
to the same data. This raises several questions:  Would a \BGLabc222 fit to the
2018 Belle measurement~\cite{Abdesselam:2018nnh} find a larger value of
$|V_{cb}|$ than that in Eq.~(\ref{belle18BGL}), closer to its inclusive determination?    
The consistency of the fitted \BGLabc122 coefficients from the 2017 and 2018 Belle measurements is only at about the $2\sigma$ level for $\tilde a_0$.

Also shown in Fig.~\ref{fig:FR12} are the fit results for the form factor ratios 
$R_{1,2}(w)$.  The \BGLabc222 fit to the tagged Belle
measurement~\cite{Abdesselam:2017kjf} indicated a substantial deviation from
heavy quark symmetry, in particular for the $R_1$ form factor
ratio~\cite{Bernlochner:2017xyx}. 
The central values, for fixed quark mass parameters, at order ${\cal O}(\lqcd/m_{c,b},\, \alpha_s)$ are~\cite{Bernlochner:2017xyx},
\begin{align}\label{R121}
R_1(1) &= 1.34 - 0.12 \eta(1) + \ldots \,, \nn\\*
R'_1(1) &= -0.15 + 0.06\, \eta(1) - 0.12\, \eta'(1) + \ldots\,,
\end{align}
where $\eta(w)$ is a ratio of a subleading and the leading Isgur-Wise function. 
With $\eta(1)$ and $\eta'(1)$ of order unity, $R_1(1)$ cannot be much below 1,
and $|R'_1(1)|$ cannot be large, without a breakdown of heavy quark symmetry.
Preliminary lattice QCD calculations~\cite{BDsLatticeAllw, Kaneko:2018mcr} also
do not indicate ${\cal O}(1)$ violations of heavy quark symmetry.
Figure~\ref{fig:FR12} shows that the \BGLabc122 fit exhibits better agreement with
heavy quark symmetry expectations for $R_1(w)$.  However, this likely arises
because $R_1(w) \propto (w+1)\, g / f$,  so setting $a_1 = 0$ constrains the
shape of the numerator. By contrast, the \BGLabc212, \BGLabc221, and \BGLabc222
fits prefer $a_1 \neq 0$, and yield $R_1(w)$ in some tension with heavy quark
symmetry and lattice QCD.

\section{Toy studies}

\begin{figure*}[t]
\includegraphics[width=0.45\textwidth]{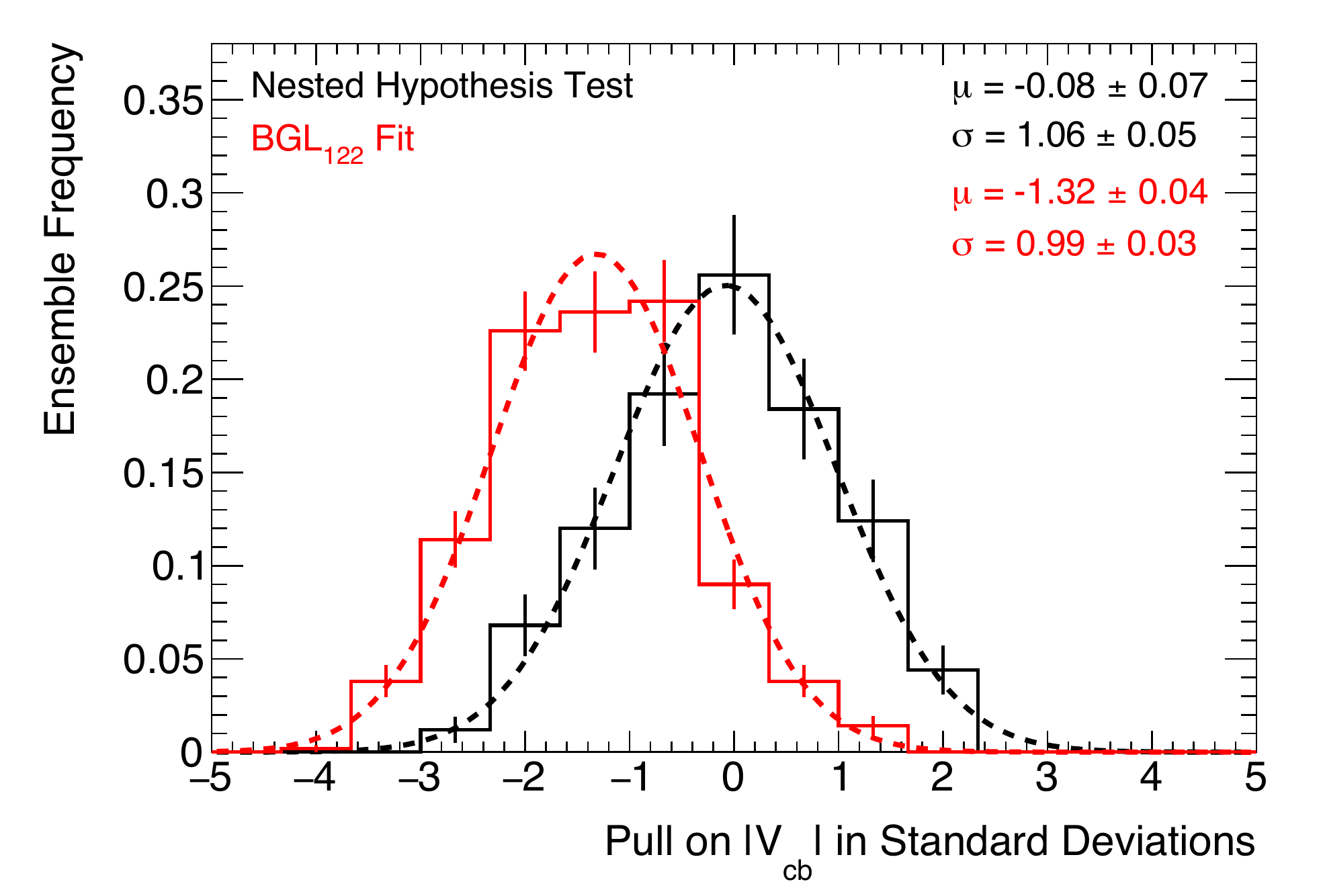} \hfil
  \includegraphics[width=0.45\textwidth]{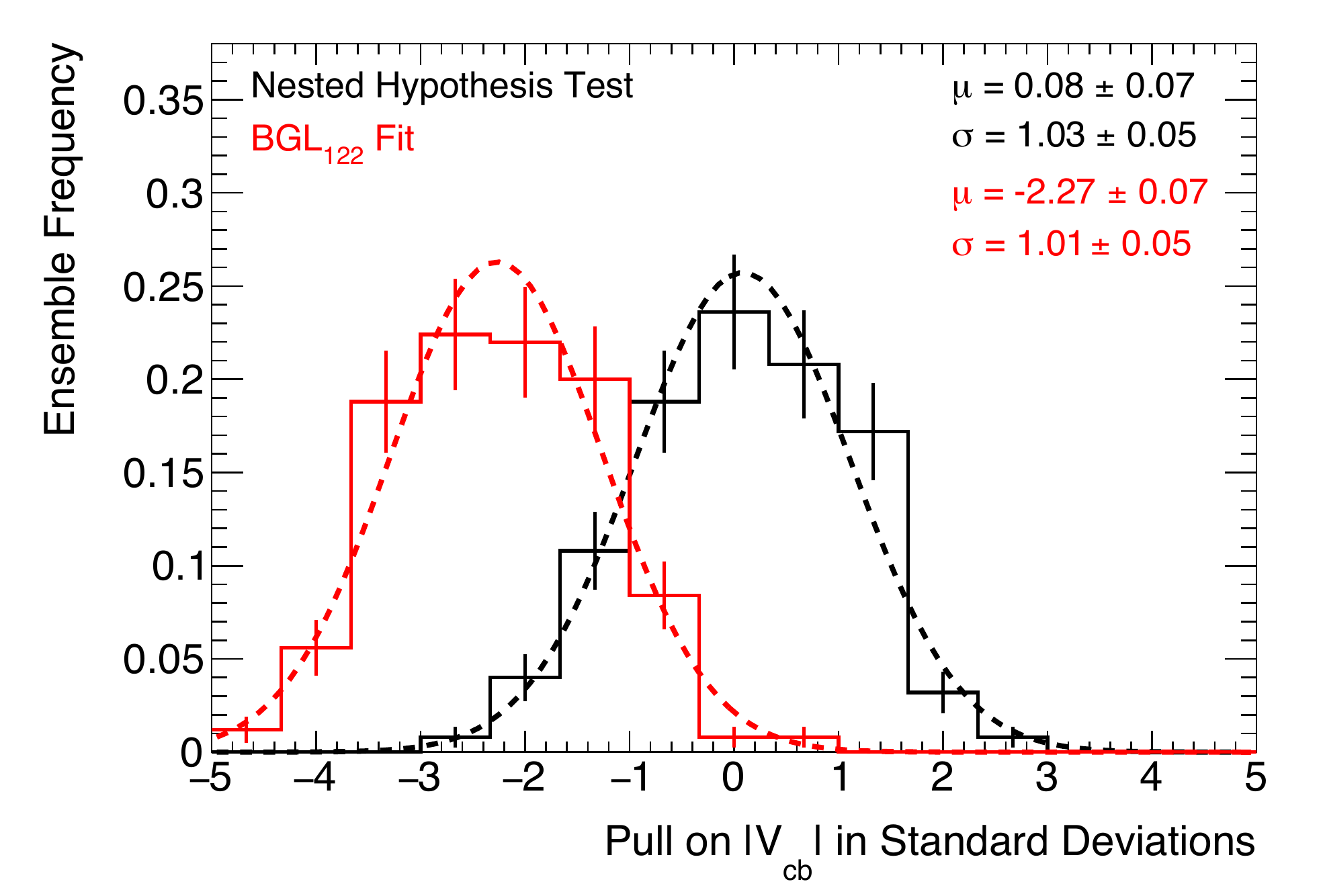} \\
\caption{The pull constructed from a large ensemble of pseudo-experiments using
3rd order terms of the $1\times$ scenario (left plot) and $10\times$ scenario
(right plot) described  in the text. The pull of the fits selected by the nested
hypothesis prescription (black) show no bias or under-coverage of uncertainties.
Also shown in red is the pull from a \BGLabc122 fit, showing a large bias on the
value of $|V_{cb}|$. Mean ($\mu$) and standard deviation ($\sigma$) from normal
distributions fitted to the ensembles are also provided.}
\label{fig:toys}
\end{figure*}

To validate the prescription outlined above, and to demonstrate that it yields
an unbiased value of $|V_{cb}|$, we carried out a toy MC study using ensembles
of pseudo-data sets.  These were generated using the \BGLabc333 parametrization,
i.e., with nine coefficients. The six lower order coefficients $\{ \tilde
a_{0,1},\, \tilde b_{0,1},\, \tilde c_{1,2}\}$ were chosen to be identical to
the \BGLabc222 fit results of Fig.~\ref{tab:chi2Vcb}. The 3rd order terms
$\{\tilde a_2,\, \tilde b_2,\, \tilde c_3\}$ were chosen according to two
different scenarios: either 1 or 10 times the size of the $\{\tilde a_1,\,
\tilde b_1,\, \tilde c_2\}$ coefficients in the \BGLabc222 fit, as shown in
Table~\ref{tab:ToyBGL}.  We call these the ``$1\times$" and ``$10\times$" scenarios,
respectively. Ensembles were constructed as follows. First, predictions for the
40 bins of the tagged measurement~\cite{Abdesselam:2017kjf} were produced.
Ensembles of pseudo-data sets were then generated using the full experimental
covariance, assuming Gaussian errors, and then each pseudo-data set was fit
according to the nested hypothesis test prescription. 

\begin{table}[b]
\begin{tabular}{c|c|c} 
\hline \hline
Parameter  &  \multicolumn{1}{c|}{$1\times$ scenario}  &
   \multicolumn{1}{c}{$10\times$ scenario}  \\
\hline
$\tilde a_2 \times 10^2$  &  $2.6954$ &  $26.954$ \\
$\tilde b_2 \times 10^2$  &  $-0.2040$ &  $-2.040$ \\
$\tilde c_3 \times 10^2$  &  $0.5350$  &  $5.350$  \\
\hline\hline
\end{tabular}
\caption{Fit coefficients used to construct the ensembles of toy experiments.
The third order terms $\{\tilde a_2, \tilde b_2, \tilde c_3\}$ are taken either
as 1 or 10 times the second order terms $\{\tilde a_1, \tilde b_1, \tilde c_2\}$
in the \BGLabc222 fit shown in Fig.~\ref{tab:BGLtag}.}
\label{tab:ToyBGL}
\end{table}

\begin{table*}[t]
\begin{tabular}{c|cccccccccccc} 
\hline \hline
& \BGLabc122 & \BGLabc212 & \BGLabc221 & \BGLabc222 & \BGLabc223 & \BGLabc232  & \BGLabc322 & \BGLabc233 & \BGLabc323 & \BGLabc332 & \BGLabc333  \\ \hline
$1\times$ scenario & 6\% & 0\% & 37\% & 27\% & 6\% & 6\%& 11\% & 0\% & 2\% & 4\% & 0.4\% \\
$10\times$ scenario & 0\% & 0\% & 8\% & 38\% & 14\% &  8\% & 16\% & 3\% & 4\% & 8\% & 1\% \\
\hline\hline
\end{tabular}
\caption{The frequency of the selected hypotheses for ensembles created with the
two scenarios for the higher order terms, as estimated with an ensemble size of
250 pseudo-data sets.}
\label{tab:frac}
\end{table*}

The frequency with which particular \BGLabc{i}{j}{k} parametrizations are
selected are shown in Table~\ref{tab:frac}, for both the $1\times$ and
$10\times$ scenarios. For each selected fit hypothesis, the recovered value,
$|V_{cb}|_{\text{rec}}$, and the associated uncertainty, $\sigma$, may then be
used to construct a pull, i.e., the normalized difference
$(|V_{cb}|_{\text{rec}} - |V_{cb}|_{\text{true}})/\sigma$,  where
$|V_{cb}|_{\text{true}}$ is the `true' value used to construct the ensembles. 
If a fit or a procedure is unbiased, the corresponding pull distribution should
follow a standard normal distribution (mean of zero, standard deviation of
unity).  In Fig.~\ref{fig:toys} the pull distributions for both the $1\times$
and $10\times$ scenarios are shown and compared to that of the \BGLabc122
parametrization.  One sees that the nested hypothesis test proposed in this
paper selects fit hypotheses that provide unbiased values for $|V_{cb}|$ in both
scenarios.  However, the \BGLabc122 fit shows significant biases. In the
ensemble tests the  \BGLabc122 fits have mean $\chi^2$ values of 41.0 and 56.6,
respectively (with 35 degrees of freedom). For the $1\times$ scenario, this
produces an acceptable fit probability on average. Nonetheless, the recovered
value of $|V_{cb}|$ is biased by about $1.3\,\sigma$.

\section{conclusions}

We studied the differences of the determinations of $|V_{cb}|$ from exclusive semileptonic $B \to D^*\ell\nu$ decays, depending on the truncation order of the BGL parametrization of the form factors used to fit the measured differential decay distributions.  
Since the 2018 untagged Belle measurement~\cite{Abdesselam:2018nnh} used a five-parameter BGL fit, 
Refs.~\cite{Grinstein:2017nlq, Bernlochner:2017xyx} used a six-parameter fit, and  Refs.~\cite{Bigi:2017njr, Jaiswal:2017rve} used an eight-parameter one, we explored differences between the five, six, seven, and eight parameter fits.

We proposed using nested hypothesis tests to determine the optimal number of fit
parameters.  For the 2017 Belle analysis~\cite{Abdesselam:2017kjf}, six parameters are preferred.  Including additional fit parameters only improves $\chi^2$ marginally.
Comparing the result of the \BGLa fit used in the 2018 untagged Belle  analysis~\cite{Abdesselam:2018nnh} to the corresponding fit to the 2017 tagged Belle measurement~\cite{Abdesselam:2017kjf},
up to $2\sigma$ differences occur, including in the values of
$|V_{cb}|$. This indicates that more precise measurements are needed to resolve tensions between various $|V_{cb}|$ determinations, and that the truncation order of the BGL expansion of the form factors has to be chosen with care, based on data.

We look forward to more precise experimental measurements, more
complete fit studies inside the experimental analysis frameworks, as well as
better understanding of the composition of the inclusive semileptonic rate as a
sum of exclusive channels~\cite{Bernlochner:2012bc, Bernlochner:2014dca}.  
Improved lattice QCD results, including finalizing
the form factor calculations in the full $w$ range~\cite{BDsLatticeAllw,
Kaneko:2018mcr} are also expected to be forthcoming.
These should all contribute to a better understanding of the determinations of $|V_{cb}|$ from exclusive and inclusive semileptonic decays, which is important for CKM fits, new physics sensitivity, $\epsilon_K$, and rare decays.

\acknowledgments

We thank Toru Iijima for organizing the KMI workshop ``Hints for New Physics in
Heavy Flavors", and thank him and Marina Artuso, Ben Grinstein, Shoji Hashimoto, Aneesh Manohar,
Sheldon Stone, and Mike Williams for useful questions and conversations.  FB was supported by the DFG Emmy-Noether Grant No.\ BE~6075/1-1.  ZL and
DR were supported in part by the U.S.\ Department of Energy under contract
DE-AC02-05CH11231.  DR was also supported in part by NSF grant
PHY-1720252.

\bibliographystyle{apsrev4-1}
\bibliography{Vcb18}

\end{document}